\documentclass{elsart}
\usepackage{amsmath,amssymb,graphicx,caption}

\begin{document}

\begin{frontmatter}



\title{A new intrinsic numerical method for PDE on surfaces \thanksref{NSC}}

\author[Chen]{Sheng-Gwo
Chen\corauthref{corl}}\ead{csg@mail.ncyu.edu.tw},
\author[Chi_Wu]{Mei-Hsiu Chi}\ead{mhchi@math.ccu.edu.tw} and
 \author[Chi_Wu]{Jyh-Yang Wu}\ead{jywu@math.ccu.edu.tw}

\address[Chen]{Department of Applied Mathematics, National Chiayi University,  Chia-Yi 600,
Taiwan.}
\address[Chi_Wu]{Department of Mathematics, National Chung Cheng
University, Chia-Yi 621, Taiwan}

\corauth[corl]{csg@mail.ncyu.edu.tw}
\thanks[NSC]{ Partially supported by NSC, Taiwan}

\begin{abstract}

In this note we shall introduce a simple, effective numerical method
for solving partial differential equations for scalar and
vector-valued data defined on surfaces. Even though we shall follow
the traditional way to approximate the regular surfaces under
consideration by triangular meshes, the key idea of our algorithm is
to develop an intrinsic and unified way to compute directly the
partial derivatives of functions defined on triangular meshes. We
shall present examples in computer graphics and image processing
applications.

\end{abstract}

\end{frontmatter}


\section{Introduction}
    Numerical approaches to solve partial differential equations (PDE's)
on surfaces have received growing interest over last decade.
However, they are still not well-understood. Partial differential
equations need to be solved intrinsically and numerically for data
defined on 3D surfaces in many applications. For instance, such
examples exist in texture synthesis (Turk\cite{Turk}, Witkin and
Kass\cite{Witkin}), vector field visualization (Diewald, Preufer and
Rumpf\cite{Clarenz}), weathering (Dorsey and Hanrahan\cite{Dorsey})
and cell-biology (Ayton, McWhirter, McMurty and Voth 2005). Usually,
surfaces are presented by triangular or polygonal forms. Partial
differential equations are then solved on these triangular or
polygonal meshes with data defined on them. The use of triangular or
polygonal meshes is very popular in all areas dealing with 3D
models. However, it has not yet been a widely accepted method to
compute differential characteristics such as principal directions,
curvatures and Laplacians (Chen and Wu\cite{Chen1}, Wu, Chen and Chi
\cite{Chen3,Chen4}, Taubin\cite{Taubin1,Taubin3}). This is because
that there is no unified, simple and effective method to compute
these first and second order differential characteristics of the
triangular or polygonal surface and to solve PDE's for data defined
on triangular or polygonal meshes. In Chen, Chi and Wu\cite{Chen5},
the authors proposed a new intrinsic simple algorithm to handle this
difficulty. In this note, we shall use this new technique to solve
PDE's on surfaces.

    In Bertalmio, Cheng, Osher and Sapiro\cite{Bertalmio} proposed a framework,
the implicit surface algorithm, to solve variational problems and
PDE's for scalar and vector-valued data defined on surfaces.Their
key idea is to use, instead of a triangular or polygonal
representation, an implicit representation. The surface under
consideration is the zero-level set of a higher dimensional
embedding function. Then they smoothly extend the original data on
the surface to the 3D domain, adapt the PDE's accordingly, and
implement all the numerical computations on the fixed Cartesian grid
corresponding to the embedding function. The advantage of their
method is the use of the Cartesian grid instead of a triangular mesh
for the numerical implementation.

\section{Our intrinsic algorithm for solving PDE's on surfaces}
    In ths section we first propose our discrete intrinsic algorithm for
solving PDE's on regular surfaces. We divide our algorithm into two
main steps: First. we approximate the given surface by a suitable
triangular mesh according to the accuracy demand. Second, we use our
new intrinsic differential method developed in Chen, Chi and
Wu\cite{Chen5} to compute the numerical PDE on the fixed triangular
mesh. The first step is now easy to implement since one can find
some good and efficient algorithms in the public domain. The
difficult part lies in the second step. Namely, how can one
effectively compute differential quantities on functions on a
triangular mesh?

  Next, we shall compare our algorithm with the implicit algorithm
proposed by Bertalmio, Cheng, Osher and Sapiro\cite{Bertalmio}. We
list the key steps of these two algorithms about solving PDE's on
surfaces as follows.

\begin{table}[h]

\begin{tabular}{|p{6cm}|p{6cm}|}
\hline  Our intrinsic algorithm & The implicit surface algorithm \\
\hline
 1.
Obtain a triangular mesh approximated to the given surface. & 1.
Obtain an implicit representation of the given fixed surface. \\

\hline
 & 2. Extend smoothly the data on the surface to the 3D
volume \\
 \hline
 & 3. Adapt PDE's accordingly \\
 \hline
 2. Use our new intrinsic differential method to compute the
numerical PDE's on the fixed triangular mesh.  &  4. Perform all the
computations on the fixed Cartesian grid corresponding to the
embedding function. \\
 \hline
\end{tabular}

\end{table}

    One can tell from this comparison that our method is much simpler
and more intrinsic. In many applications, one usually starts with
triangular meshes instead of regular surfaces. In this case, we do
not need Step 1 in our intrinsic algorithm. However, in the implicit
surface algorithm, one will need one extra processing step:
Construct an accurate implicit surface from a triangular mesh. Note
that the triangular mesh may compose of a lot of triangles. This
will cost large computations to obtain the accurate implicit
surface.

    Next, we shall describe a new, simple and effective method
to define the discrete gradient and the discrete LB operator on
functions on a triangular mesh. In order to do so, we first recall
the gradient and the LB operator on functions in a regular surface
$\Sigma$ in the 3D Euclidean space $\mathbb{R}^3$.

\subsection{Gradient and LB operators on regular surfaces }
We consider a parameterization $x : U \rightarrow \Sigma$ at a point
$p$, where $U$ is an open subset of the 2D Euclidean space
$\mathbb{R}^2$. We can choose, at each point $q$ of $x(U)$, a unit
normal vector $N(q)$. The map $N: x(U) \rightarrow S^2$ is the local
Gauss map from an open subset of the regular surface $\Sigma$ to the
unit sphere in the 3D Euclidean space $\mathbb{R}^3$. The Gauss map
$N$ is differentiable. Denote the tangent space of $\Sigma$ at the
point $p$ by $T\Sigma _p = \{ v \in \mathbb{R}^3 : v \bot N(p) \}$.
The tangent space $T\Sigma _p$ is a linear space spanned by $\{ x_u,
x_v \}$ where $u,v$ are coordinates for $U$.

The gradient $\nabla g$ of a smooth function $g$ on $\Sigma$ can be
computed from

\begin{equation}\label{ nabla_g}
\nabla g = \frac{g_uG - g_vF}{EG-F^2}x_u+
\frac{g_vE-g_uF}{EG-F^2}x_v
\end{equation}

where $E$, $F$, and $G$ are the coefficients of the first
fundamental form and

\begin{equation}\label{g_u_g_v}
\left \{ \begin{array}{ll}
 g_u & = \frac{\partial g(x(u,v))}{\partial u} \cr
 g_v & = \frac{\partial g(x(u,v))}{\partial v} \end{array} \right.
\end{equation}
See do Carmo\cite{Do} for the details. Note that the gradient
$\nabla g$ assigns to each point $q$ in $\Sigma$ a tangent vector
$\nabla g(q)$ such that we have for all $v \in T\Sigma _q$,
\begin{equation}\label{dot_gradient_g_v}
\langle \nabla g(q), v \rangle _q = \frac{dg(\gamma (t) )}{dt}\|
_{t=0}
\end{equation}
where the smooth curve $\gamma (t)$ is in $\Sigma$ with $\gamma (0)$
and $\gamma ' (0) = v$.

    The LB operator $\triangle$ acting on the function $g$ is defined by the
integral duality

\begin{equation}
( \triangle g, \phi ) = - ( \nabla g, \nabla \phi)
\end{equation}
for all smooth function $\phi$ on $\Sigma$. A direct computation
yields the following local representation for the LB operator on a
smooth function $g$:

\begin{equation}
\begin{array}{ll}
\nabla g & = \frac{1}{\sqrt{EG-F^2}}\left [ \frac{\partial}{\partial
u}(\frac{G}{\sqrt{EG-F^2}}\frac{\partial g}{\partial u}) -
\frac{\partial }{\partial u}(\frac{F}{\sqrt{EG-F^2}}\frac{\partial
g}{\partial v}) \right ] \cr
 & +\frac{1}{\sqrt{EG-F^2}}\left [ \frac{\partial}{\partial
v}(\frac{E}{\sqrt{EG-F^2}}\frac{\partial g}{\partial v}) -
\frac{\partial }{\partial v}(\frac{F}{\sqrt{EG-F^2}}\frac{\partial
g}{\partial u}) \right ]
\end{array}
\end{equation}

To move from regular surfaces to triangular meshes, one need to
avoid the problem of local parametrization $x$ around a vertex $p$.
In other word, one does not have the fist fundamental form $E$, $F$,
$G$ and their derivatives for the computation of the gradient and
the Laplacian operator of a function on a triangular mesh. To handle
this problem, we give a novel method in Chen, Chi and Wu\cite{Chen5}
to compute these differential quantities. The primary ideas were
developed in Chen, Chi and Wu\cite{Chen4} where we try to estimate
the discrete partial derivatives for 2D scattered data points.

\subsection{ A new discrete algorithm: local tangential lifting(LTL) method}

In this section we shall describe a unified, simple and effective
method to define the discrete gradient and the discrete Laplacian
operator on functions on a triangular mesh. The primary ideas were
developed in Chen, Chi and Wu\cite{Chen3,Chen4} where we try to
estimate the discrete partial derivatives of functions on 2D
scattered data points. Indeed, the method that we shall use to
develop our algorithm is divided into two main steps: first we lift
the 1-neighborhood points to the tangent space and obtain a local
tangential polygon. Second, we use some geometric idea to lift
functions and vectors to the tangent space and then we can compute
their derivatives in the 2D tangent space. This means that the
lifting process allows us to reduce the 2D curved surface problem to
the 2D Euclidean problem and hence the methods in \cite{Chen4}
and\cite{Hiroshi} can be applied.

\begin{figure}
{\center
\includegraphics[height=6.2cm]{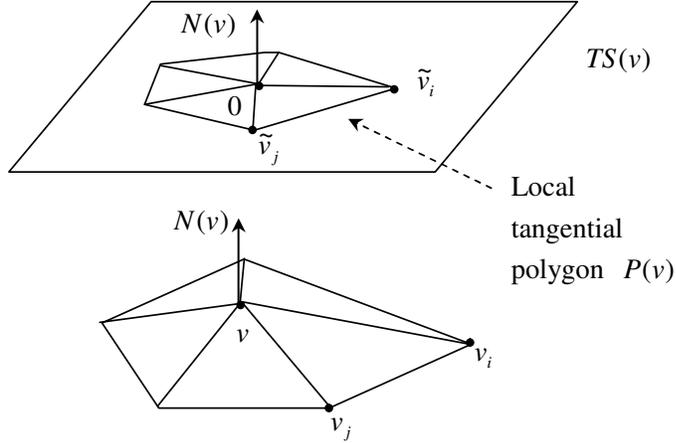}
\caption{ The tangential polygon $P(v)$.} \label{tangential}
 }
\end{figure}

Consider a triangular mesh $S=(V,F)$, where $V=\{ v_i | 1 \leq i
\leq n_V \}$ is the list vertices and $F = \{ f_k | 1 \leq k \leq
n_F \}$ is the list of triangles. Next, we introduce the notion  of
the local tangential polygon $P(v)$ at the vertex $v$  of $S$ as
follows:
\begin{enumerate}
\item The normal vector $N(v)$ at the vertex $v$ in $S$ is given by
\begin{equation}\label{N_v}
N(v) = \frac{\sum_{f \in T(v)} \omega_fN_f}{\left \|\sum_{f \in
T(v)} \omega_fN_f \right \|}
\end{equation}
where $N_f$ is the unit normal to a triangle face $f$ and the
centroid weight is given in \cite{Chen1} by
\begin{equation}\label{omega_f}
\omega_f = \frac{\frac{1}{\| G_f - v \|^2}}{\sum_{\tilde{f} \in
T(v)}\frac{1}{\| G_{\tilde{f}} - v \|^2}}.
\end{equation}
Here, $G_f$ is the centroid of the triangle face $f$ determined by
\begin{equation}G_f = \frac{v_i+v_j+v}{3}.\end{equation}
\item The tangent plane $TS(v)$ of $S$ at $v$ is now determined by
\begin{equation}TS(v) = \{ w \in \mathbb{R}^3 | w \bot N(v)
\}.\end{equation}
\item The local tangential polygon $P(v)$ of $v$ in $TS(v)$ is formed by the vertices $\tilde{v}_i$ which is the lifting vertex of $v_i$ adjacent to $v$ in $S$.
\begin{equation} \tilde{v}_i = ( v_i-v) - < v_i - v, N(v)>N(v).\end{equation} as in figure
\ref{tangential}.
\end{enumerate}

Let $h$ be a function on $V$. We will lift locally the function $h$
to a function, denoted by $\tilde{h}_v$, on the vertices
$\tilde{v}_i$ in $P(v)$ by simply setting
\begin{equation}
\tilde{h}_v(\tilde{v}_i) = h(v_i).
\end{equation} And $\tilde{h}_v(\vec{0}) = h(v)$ where $\vec{0}$ is the origin of
$TS(v)$. One can then extend the function $\tilde{h}_v$ to a
piecewise linear function, still denoted by $\tilde{h}_v$, on $P(v)$
as follows.

Consider a face $f$ with vertices $v$, $v_i$ and $v_j$ in $F$. We
obtain a lifting face $\tilde{f}$ with vertices $\vec{0}$,
$\tilde{v}_i$ and $\tilde{v_j}$  in $P(v)$. Every point  $p$ in
$\tilde{f}$ can be written as a linear combination of $\tilde{v}_i$
and $\tilde{v_j}$ . That is, $ p = a\tilde{v}_i+b\tilde{v}_j$ where
$a, b \geq 0$ and $a+b \leq 1$. Then we define
\begin{equation}
\tilde{h}_v(p) = a \tilde{h}_v(\tilde{v}_i) + b
\tilde{h}_v(\tilde{v}_j) +( 1-a-b)\tilde{h}_v(\vec{0}).
\end{equation}
Hence, the extended function $\tilde{h}_v$ is affine on each
triangle $\tilde{f}$ of $P(v)$ and is differentiable on $\tilde{f}$.
The gradient $\nabla (\tilde{h}_v)_{\tilde{f}}$ of $\tilde{h}_v$ at
the origin $\vec{0}$ can be obtained by \begin{equation} \nabla
(\tilde{h}_v)_{\tilde{f}}(\vec{0}) = \alpha \tilde{v}_i + \beta
\tilde{v}_j\end{equation} where the coefficients $\alpha$ and
$\beta$ satisfy the relations:
\begin{equation} \left\{
\begin{array}{ll} \tilde{h}(\tilde{v}_i) - \tilde{h}_v(\vec{0}) & =<
(\nabla \tilde{h}_v)_{\tilde{f}}(\vec{0}), v_i> \cr
\tilde{h}(\tilde{v}_j) - \tilde{h}_v(\vec{0}) & = < (\nabla
\tilde{h}_v)_{\tilde{f}}(\vec{0}),v_j>
\end{array}\right.\end{equation}

As easy computation gives
\begin{equation}\begin{pmatrix} \alpha \cr
\beta
\end{pmatrix} = \begin{pmatrix} <\tilde{v}_i,\tilde{v}_i> &
<\tilde{v}_i,\tilde{v}_j> \cr <\tilde{v}_i,\tilde{v}_j> &
<\tilde{v}_j, \tilde{v}_j> \end{pmatrix}^{-1} \begin{pmatrix}
\tilde{h}_v(\tilde{v}_i) - \tilde{h}_v(\vec{0}) \cr
\tilde{h}_v(\tilde{v}_j) - \tilde{h}_v(\vec{0}).
\end{pmatrix}\end{equation}

To obtain the gradient $\nabla h(v)$ of $h$ on $S$ at the vertex
$v$, we use again the weighted combination method. Namely, we set
\begin{equation}\nabla h(v) = ( \nabla \tilde{h}_v)(\vec{0}) = \sum_{\tilde{f}
\in P(v)} \omega_{\tilde{f}}(\nabla
\tilde{h}_v)_{\tilde{f}}(\vec{0})\end{equation} with
\begin{equation}\label{centroidweight}
\omega_{\tilde{f}} = \frac{\frac{1}{\| G_{\tilde{f}}\|^2}}{\sum_{f
\in P(v)} \frac{1}{\| G_{f}\|^2}} \end{equation}
 where $G_{\tilde{f}}$ is the centroid of the lifting triangle face
$\tilde{f}$ and is determined by
\begin{equation}G_{\tilde{f}} =
\frac{\tilde{v}_i+\tilde{v}_j}{3}.\end{equation}

\begin{figure}
{\center
\includegraphics[height=6.2cm]{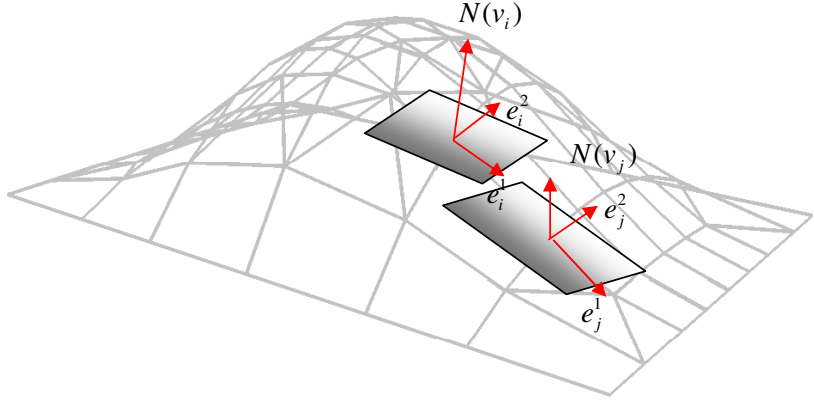}
\caption{parallel transport}\label{parallel}
 }
\end{figure}

Next we explain how to obtain a good discrete Laplacian $\triangle
h(v)$ of a function $h$ on the triangular mesh $S$. From the
discussions above, we obtain the gradient $\nabla h(v)$ of $h$ on
$S$ at each vertex $v$. We can use the method of parallel transport
to lift the vector $\nabla h(v_i)$ at $v_i$ to a vector $\nabla
\tilde{ h}(\tilde{v}_i)$ in the tangential space $TS(v)$. The idea
is to define a orthonormal linear map from $TS(v_i)$ to $TS(v)$. To
do so, we choose an orthonormal basis for $TS(v)$ by
\begin{equation} \left \{\begin{array}{l} N(v) \cr e_1 =
\frac{(v_i-v) - <v_i-v,N(v)>N(v)}{\|(v_i-v) - <v_i-v,N(v)>N(v)\|}
\cr e_2 = N(v) \times e_1
\end{array}\right. .\end{equation}
The corresponding orthonormal basis for $TS(v_i)$ is then given by
\begin{equation}
\left \{\begin{array}{l} N(v_i) \cr \tilde{e}_1 = \frac{(v_i-v) -
<v_i-v, N(v_i)>N(v_i)}{ \|(v_i-v) - <v_i-v, N(v_i)>N(v_i) \|} \cr
\tilde{e}_2 = N(v_i) \times \tilde{e}_1
\end{array}\right. .\end{equation}
Then, the linear map $L$ of the parallel transport is given by
\begin{equation} L(w) = a e_1 + b e_2 \in TS(v) \end{equation}
for $w = a\tilde{e}_1 + b\tilde{e}_2 $ in $TS(v_i)$. See figure
\ref{parallel}.

\begin{figure}
{\center
\includegraphics[height=6.2cm]{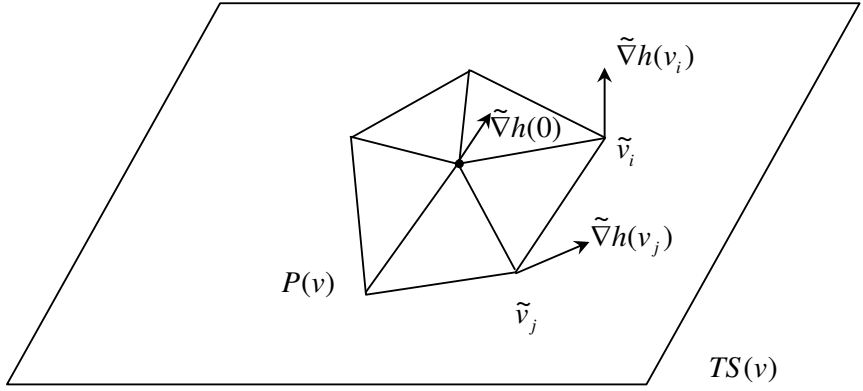}
\caption{$\nabla \tilde{h}$ }\label{tildeh}
 }
\end{figure}

In this way, we can set the tangential gradient $\nabla\tilde{ h}$
at $\tilde{v}_i$ by
\begin{equation} \nabla \tilde{h}(\tilde{v}_i) =
L(\nabla h (v_i) ). \end{equation}
 Hence we obtain a tangential
gradient $\nabla\tilde{ h}$ of $h$ at each vertex $\tilde{v}_i$ in
the tangential polygon $P(v)$ and we also set
\begin{equation}
\nabla \tilde{ h}(\vec{0}) = \nabla h(v). \end{equation}
 See figure \ref{tildeh}.

Fix an orthonormal basis $\{ e_1, e_2 \}$ for $TS(v)$. The
tangential gradient $\nabla \tilde{h}$ can be written as
\begin{equation} \nabla \tilde{h}(\tilde{v}_i) = a(\tilde{v}_i)e_1 +
b(\tilde{v_i})e_2. \end{equation}
 The coefficients $a(\tilde{v}_i)$
and $b(\tilde{v}_i)$ can now be viewed as functions on the vertices
$\tilde{v}_i$ of the tangential polygon $P(v)$. As before, we can
then obtain their gradients $(\nabla a)(\vec{0})$ and $(\nabla
b)(\vec{0})$ at origin. Namely,

\begin{equation}
\begin{array}{ll} (\nabla a)(\vec{0}) & = \sum_{\tilde{f} \in
P(v)} \omega_{\tilde{f}} (\nabla a)_{\tilde{f}}(\vec{0}) \cr (\nabla
b)(\vec{0}) & = \sum_{\tilde{f} \in P(v)} \omega_{\tilde{f}} (\nabla
b)_{\tilde{f}}(\vec{0})
\end{array}\end{equation}
with the centroid weights $\omega _{\tilde{f}}$ as in
(\ref{centroidweight}).

Put them in the matrix form to give
\begin{equation}
\begin{array}{ll} (\nabla a)(\vec{0}) & = a_{11}e_1
+ a_{21}e_2 \cr (\nabla b)(\vec{0}) & = a_{12}e_1 + a_{22} e_2.
\end{array}
\end{equation}
 Therefore we have the Laplacian $\triangle h(v)$ of $h$ at the
vertex $v$ of $S$ :
\begin{equation}
\triangle h(v) = a_{11}+ a_{22}.
\end{equation}

Theoretically the definition of the Laplacian $\triangle h(v)$ is
independent of the choice of the orthonormal basis $\{ e_1, e_2 \}$.

\section{Linear diffusion}

As a simple example to illustrate our new algorithm, let us consider
a linear diffusion equation on a regular surface $\Sigma$:

\begin{equation}\label{heatequation}
u_t - \triangle u = g \mbox{ ~~ on ~~ } \Sigma \times I
\end{equation}
for  $u : \Sigma \times I \rightarrow \mathbb{R} $, $I \subset
\mathbb{R}$, where  $\triangle$ is the surface Laplacian on $\Sigma$
and $g : \Sigma \times I \rightarrow \mathbb{R}$ is a smooth
function on $\Sigma$. For the numerical implementation of our
intrinsic algorithm, we take the regular surface $\Sigma$ to be (1)
the unit sphere $S^2$ or (2) a torus $T^2$. In the case of the
sphere $S^2$, we consider the function

\begin{equation}\label{function_g}
g(x) = x_1 \mbox{ ~~ for ~~ } x = (x_1,x_2,x_3) \in S^2.
\end{equation}

\begin{figure}
{\center
\includegraphics[scale = 1]{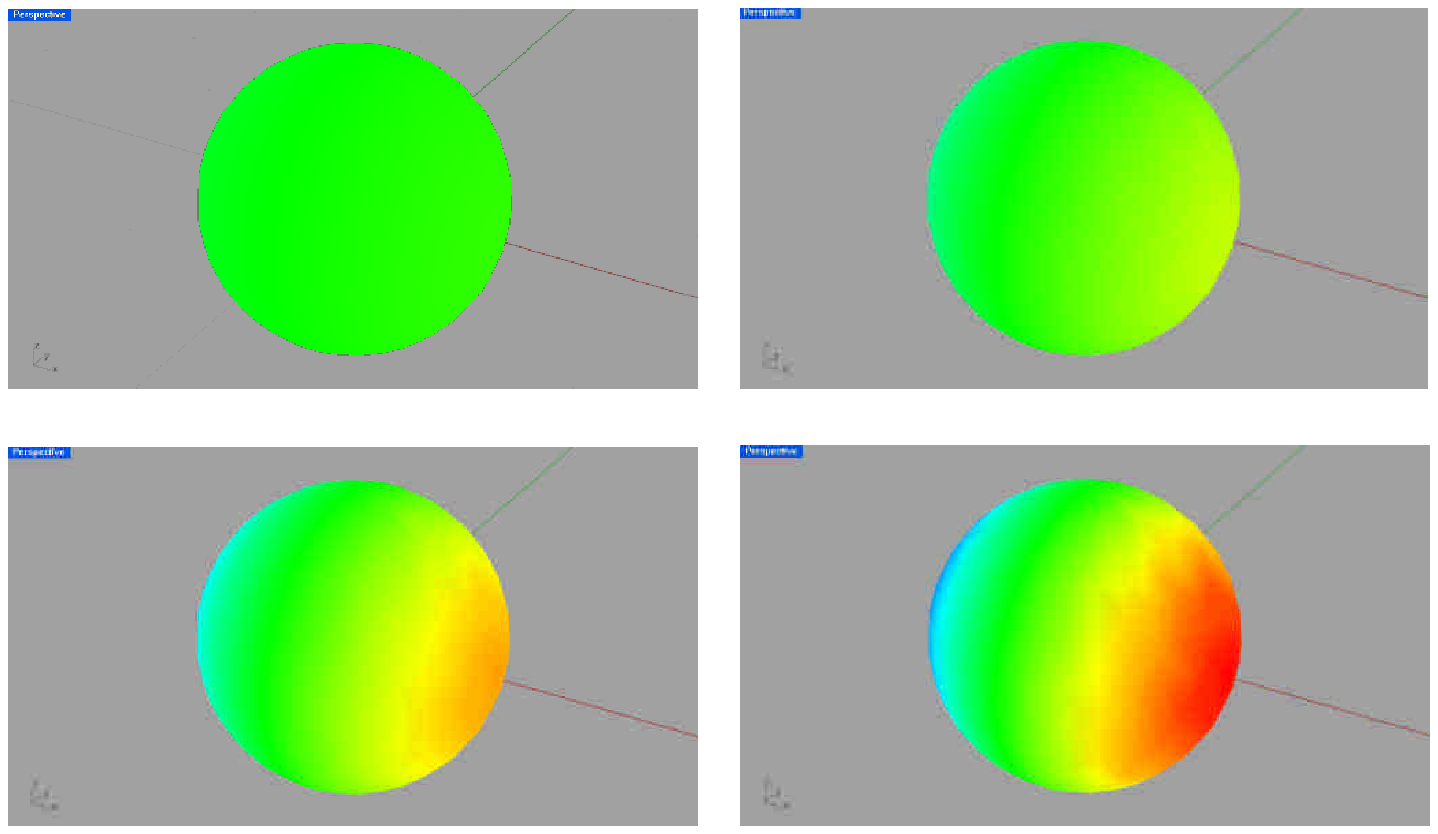}
\caption{sphere}\label{sphere}
 }
\end{figure}

\begin{figure}
{\center
\includegraphics[scale = 1]{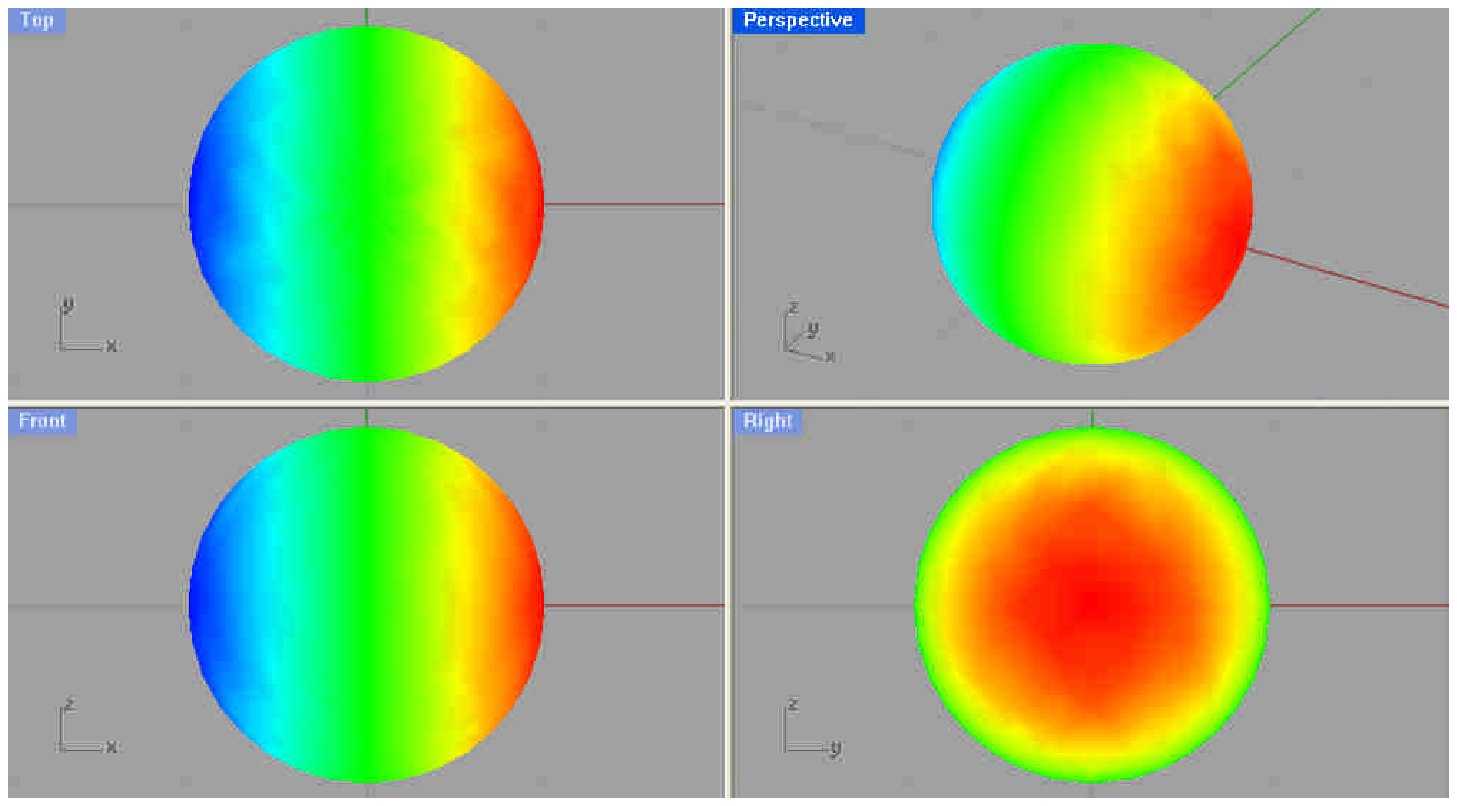}
\caption{sphere (stationary solution) }\label{shpere_final}
 }
\end{figure}

Figure \ref{sphere} and figure \ref{shpere_final} give the solution
of (\ref{heatequation}) and (\ref{function_g}) with initial
functions $u(x,0) = 0$. Different time steps are shown until the
stationary solution is reached.

Consider  the torus $T^2_{(a,r)} = ( ( a+r \cos x)\cos y, (a+r \cos
x)\sin y, \sin x)$ for $x,y \in [ 0, 2\pi]$ with $a > r > 0$. we
take $ a=2$, $r=1$ and choose the function

\begin{equation}\label{function_g2}
g(x,y) = x  \mbox{ ~~ for ~~ } x,y  \in [0, 2 \pi]
\end{equation}

\begin{figure}
{\center
\includegraphics[scale = 1]{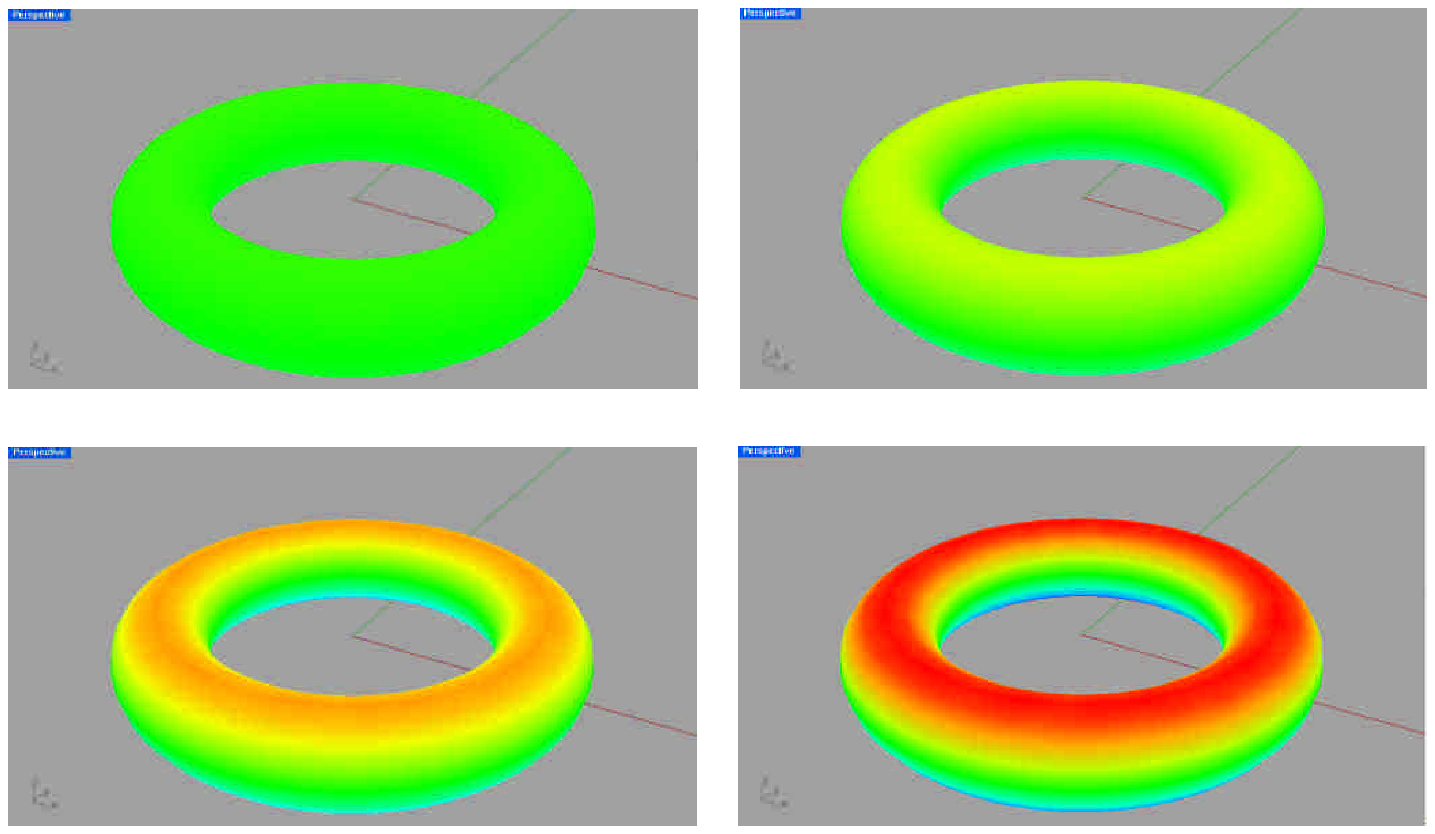}
\caption{torus}\label{torus}
 }
\end{figure}

\begin{figure}
{\center
\includegraphics[scale = 1]{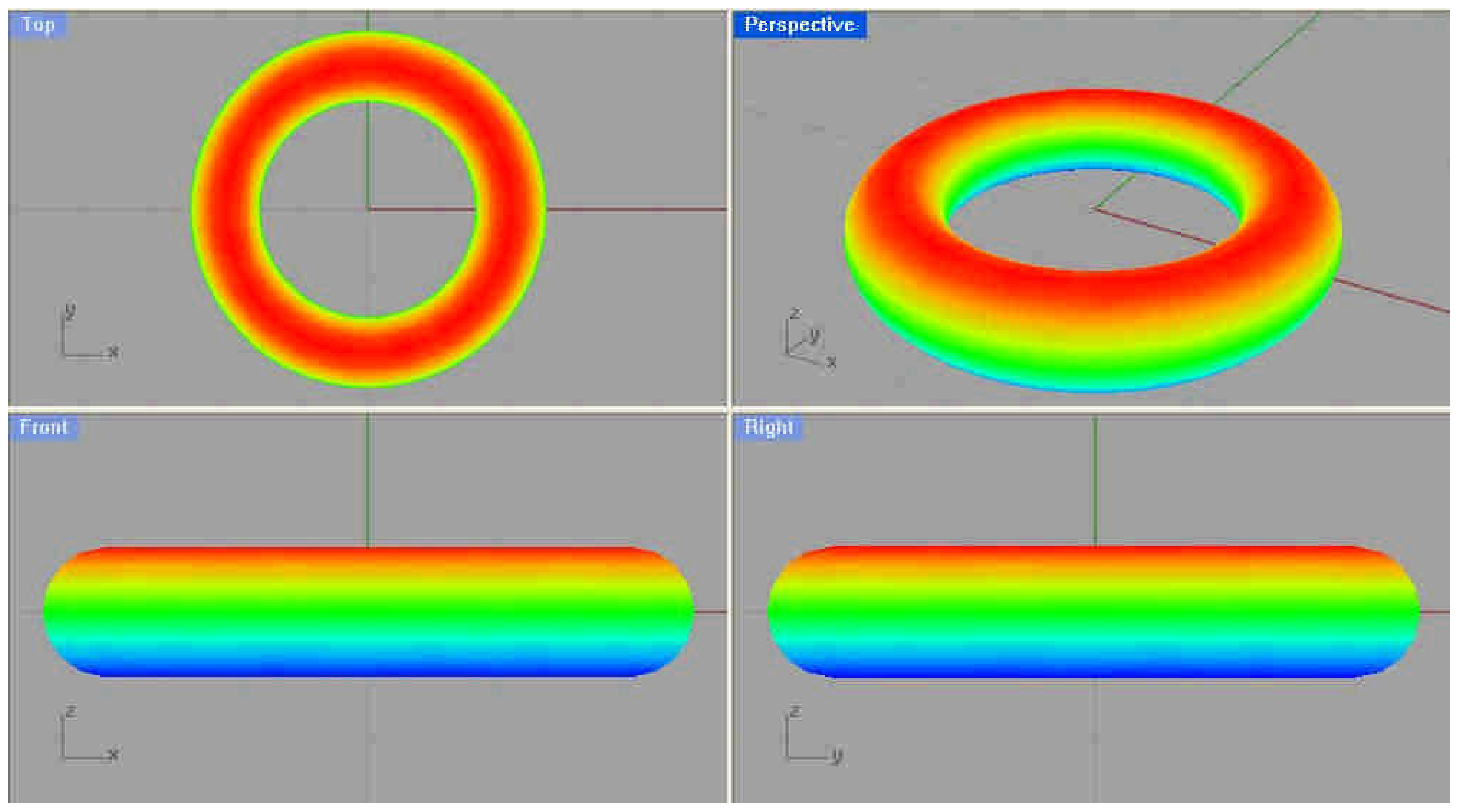}
\caption{torus (stationary solution)}\label{torus_final}
 }
\end{figure}

Figure \ref{torus} and figure \ref{torus_final} give the solution of
(\ref{heatequation}) and (\ref{function_g2}) with initial functions
$u(x,0) = 0$. As above, different timesteps are depicted until the
stationary solution is reached.

\section{Reaction-diffusion textures}

The original idea about how reaction-diffusion equations can be used
to create patterns was first introduced in (Turing\cite{Turing}).
The basic idea is to have a number of chemicals that diffuse at
different rates and that react with each others. After the works of
(Turk\cite{Turk}, Witkin and Kass\cite{Witkin}), the use of
reaction-diffusion equations for texture synthesis attracted a lot
of attentions in computer graphics. Turk , Witkin and Kass used
these equations for planar textures and textures on surfaces. Then
the patterns are analyzed by assigning a brightness value to the
concentration of one of the "chemicals".

Consider two chemicals $u_1$ and $u_2$ on a surface $\Sigma$. In a
simple isotropic model, we have

\begin{equation}\label{Du} \left \{\begin{array}{ll} \frac{\partial u_1}{\partial t} &=f(u_1,u_2)
+ \alpha \triangle u_1 \cr \frac{\partial u_2}{\partial t}
&=g(u_1,u_2) + \beta \triangle u_2\end{array} \right. \end{equation}

where $\alpha$ and $\beta$ are two constants representing the
diffusion rates and $f$ and $g$ are functions that describe the
reaction. For simple isotropic patterns, Turing chose the functions
$f$ and $g$ to be

\begin{equation}\label{fg} \left \{ \begin{array}{ll}
f(u_1,u_2) &= s(16-u_1u_2) \cr g(u_1,u_2) & = s(u_1y_2-y_2-\gamma)
\end{array}\right. \end{equation}
where $s$ is a constant and $\gamma$ is a random function giving the
irregularities in the chemical concentration.

    By using our intrinsic method described in the previous section, we
can easily generate textures on surfaces without the elaborated
schemes employed in (Turk 1991, Witkin and Kass 1991).

\begin{figure}
{\center
\includegraphics[scale = 1]{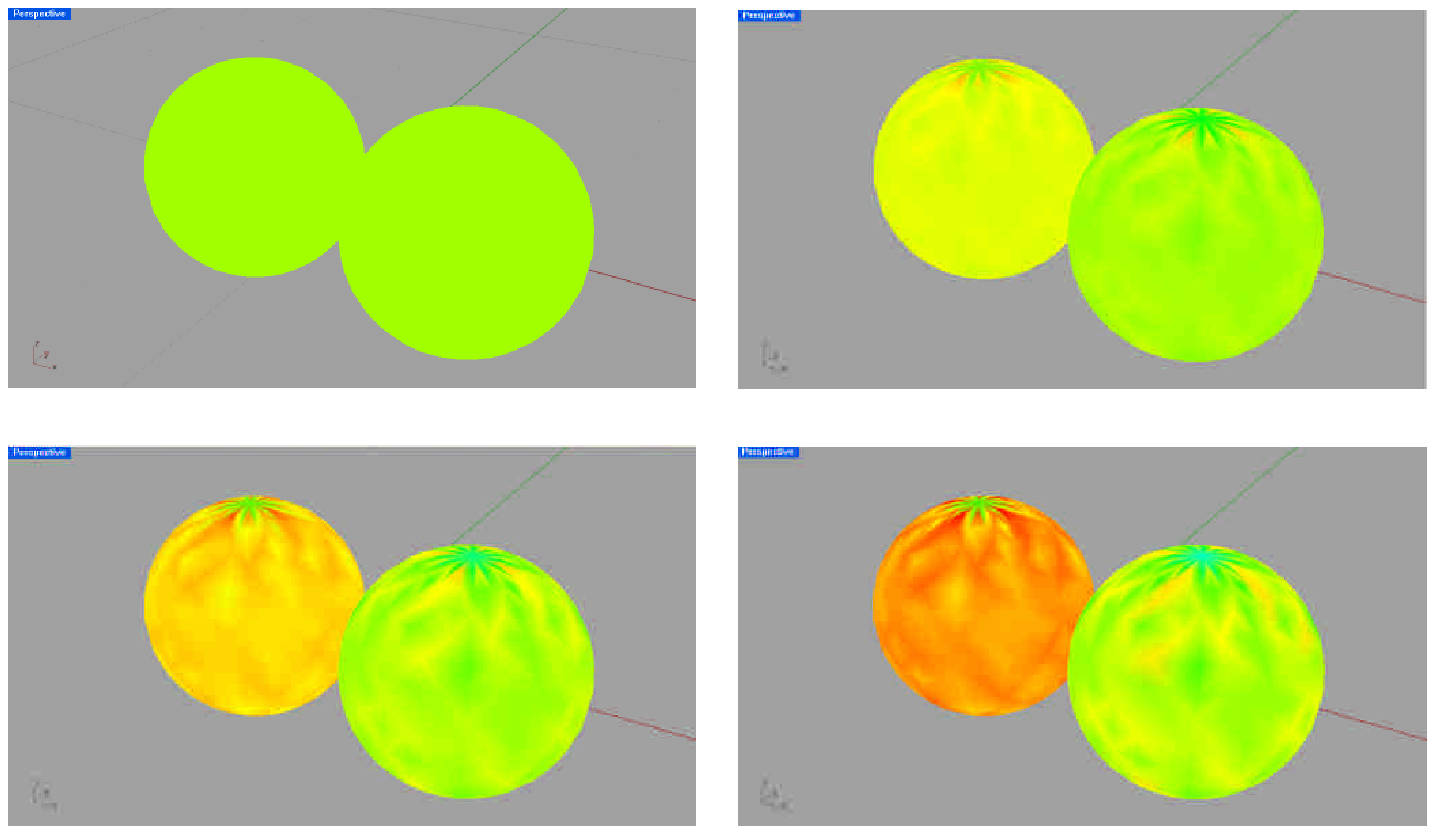}
\caption{sphere}\label{diffusion_sphere}
 }
\end{figure}

\begin{figure}
{\center
\includegraphics[scale = 1]{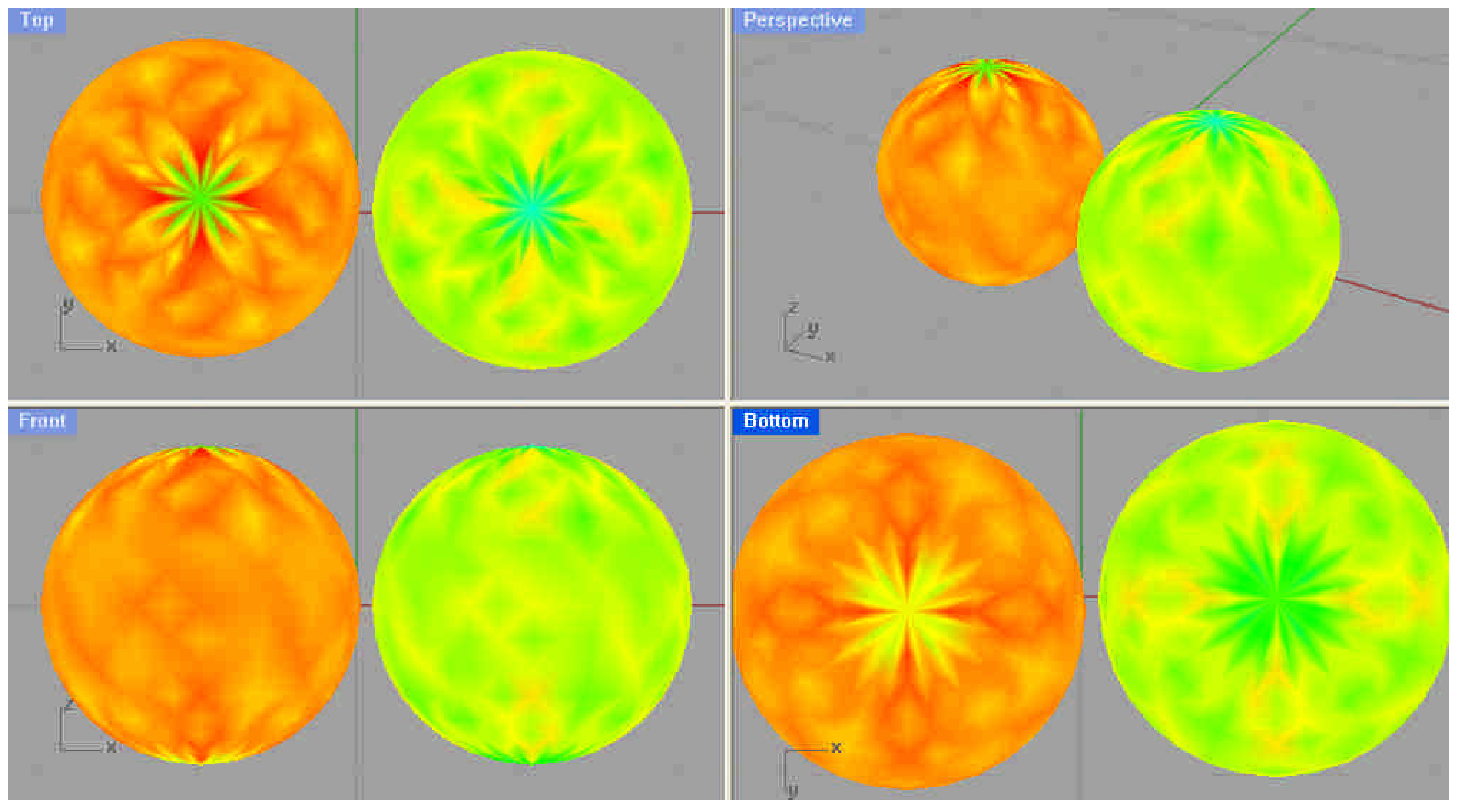}
\caption{sphere (stationary solution) }
\label{diffusion_sphere_final}
 }
\end{figure}

In the case of the sphere $S^2$, figure \ref{diffusion_sphere} and
figure \ref{diffusion_sphere_final} give the solution of (\ref{Du})
and (\ref{fg}) with initial functions and Different timesteps are
shown until the stationary solution is reached.

\begin{figure}
{\center
\includegraphics[scale = 1]{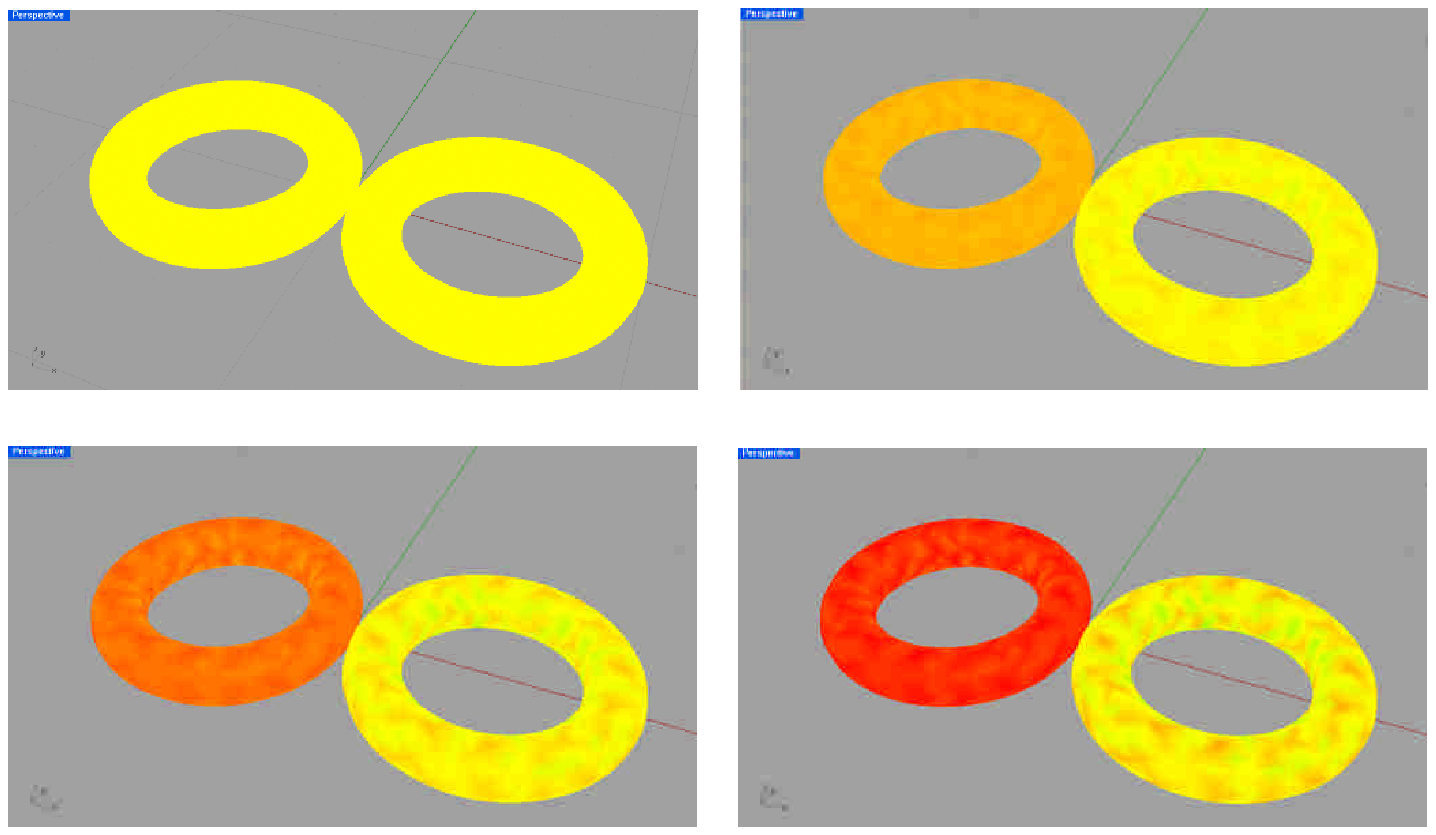}
\caption{torus}\label{diffusion_torus}
 }
\end{figure}

\begin{figure}
{\center
\includegraphics[scale = 1]{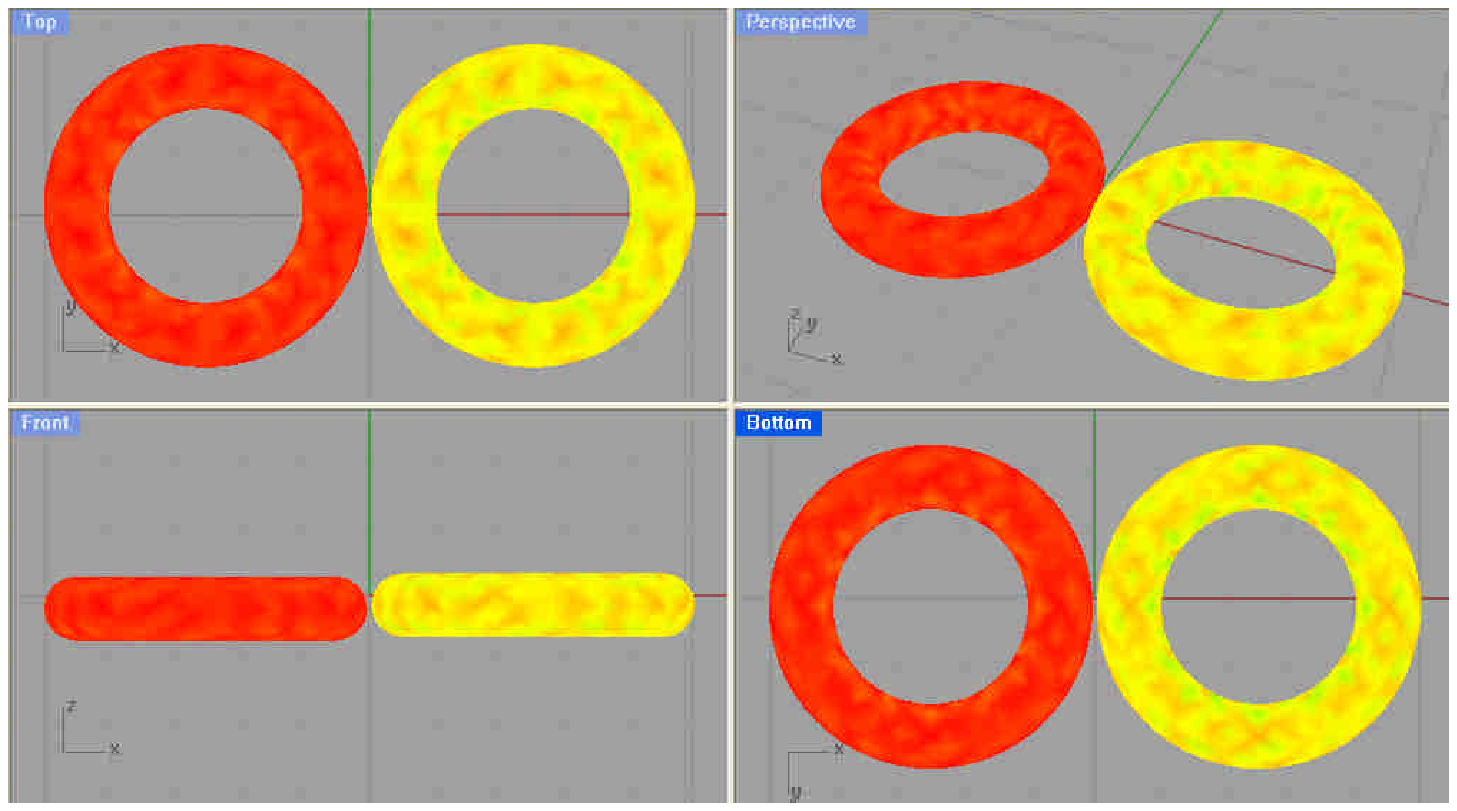}
\caption{torus (stationary solution)}\label{diffusion_torus_final}
 }
\end{figure}

On the torus $T^2_{(a,r)} = ( ( a+r \cos x)\cos y, (a+r \cos x)\sin
y, \sin x)$ for $x,y \in [ 0, 2\pi]$ figure \ref{diffusion_torus}
and figure \ref{diffusion_torus_final} give the solution of
(\ref{Du}) and (\ref{fg}) with initial functions $u_1(x,0) =
u_2(x,0) = 1$ and $ \alpha = 1$, $\beta = s = 2$, $\gamma = 0$.
Different timesteps are shown until the stationary solution is
reached.

\end{document}